\documentclass[reprint, %linenumbers,
 amsmath,amssymb,
 aps, physrev,
]{revtex4-2}
\usepackage[english]{babel}

\usepackage[letterpaper,top=2cm,bottom=2cm,left=3cm,right=3cm,marginparwidth=1.75cm]{geometry}

\usepackage{floatrow}
\usepackage[dvipsnames]{xcolor}
\usepackage{graphicx}
\usepackage[colorlinks=true, allcolors=blue]{hyperref}

\usepackage{orcidlink}

\begin{document}

%\preprint{APS/123-QED}

%\title{Multiplexed dual-comb spectroscopy of complex systems}
\title{Enhanced multichannel dual-comb spectroscopy of complex systems}

\author{Razmik Aramyan\,\orcidlink{0009-0007-2773-7343}}
 \email[Contact author: ] {aramyanr@uni-mainz.de}
\affiliation{
 Johannes Gutenberg-Universit\"at Mainz, 55128 Mainz, Germany
}%
\affiliation{
 Helmholtz-Institut Mainz, GSI Helmholtzzentrum f\"ur Schwerionenforschung, 55128 Mainz, Germany
}%
\author{Oleg Tretiak\,\orcidlink{0000-0002-7667-2933}}
\affiliation{
 Johannes Gutenberg-Universit\"at Mainz, 55128 Mainz, Germany
}%
\affiliation{
 Helmholtz-Institut Mainz, GSI Helmholtzzentrum f\"ur Schwerionenforschung, 55128 Mainz, Germany
}%
\author{Sushree S. Sahoo\,\orcidlink{0000-0001-9431-4075}}
\affiliation{
 Johannes Gutenberg-Universit\"at Mainz, 55128 Mainz, Germany
}%
\affiliation{
  Helmholtz-Institut Mainz, GSI Helmholtzzentrum f\"ur Schwerionenforschung, 55128 Mainz, Germany
}%
\author{Dmitry Budker\,\orcidlink{0000-0002-7356-4814}}
\affiliation{%
  Johannes Gutenberg-Universit\"at Mainz, 55128 Mainz, Germany
}%
\affiliation{
 Helmholtz-Institut Mainz, GSI Helmholtzzentrum f\"ur Schwerionenforschung, 55128 Mainz, Germany
}%
\affiliation{
 Department of Physics, University of California, Berkeley, CA 94720, USA
}%

\date{\today}% It is always \today, today,

\begin{abstract}
A multichannel dual-comb spectroscopy (DCS) approach for high-resolution, broadband spectral measurements is presented, demonstrating its effectiveness in studying complex atomic systems. By implementing a photodetector array, we enhance DCS capabilities, addressing the fundamental trade-off between signal-to-noise ratio (SNR) and spectral coverage. 
To resolve ambiguities in frequency conversion, we introduced a relative offset in beat note frequency, ensuring accurate spectral reconstruction. As a proof of concept, the absorption spectrum of samarium (Sm) vapor is investigated over a 52\,nm range, and several previously unreported absorption lines are detected. This is a step toward ``Spectroscopy 2.0”, enabling massively parallel spectroscopic measurements (including those at $>$100\,T magnetic fields) crucial for atomic physics and fundamental interactions research. 
\end{abstract}

\maketitle

\section{Introduction}
A diverse selection of atomic systems with a range of properties is essential for optimizing sensitivity to specific interactions of interest in fundamental physics experiments. This need is particularly pressing in the rapidly expanding field of searches for new physics using atoms and molecules \cite{Safronova|2018|Search_for_new_physics}. While atomic databases provide valuable spectral information \cite{Kramida1999}, their coverage remains incomplete, especially for rare-earth and actinide elements. The primary reason for these gaps in atomic tables is the complexity of dense atomic spectra, which present significant challenges for both experimental measurements and theoretical modeling.
In ref.\,\cite{Battesti|2018|Spectroscopy2.0_overview}, a novel approach to spectroscopy (called Spectroscopy 2.0) is introduced, integrating three key elements: (1) massively parallel spectroscopic tool, (2) tunable high magnetic fields that introduce a new spectroscopic dimension, and (3) advanced many-body theory enhanced by modern computational power and machine learning.

In this paper, we take the first step toward implementing Spectroscopy 2.0: development of a massively parallel spectroscopic tool. As the core technique, we employ dual-comb spectroscopy (DCS), a revolutionary method renowned for its high-resolution, high-sensitivity, and broadband spectral measurements at remarkable speed \cite{Hansch|2006|Nobel_lecture,Schiller2002, Picque|2019|Frequency_comb_spectroscopy_review, Coddington|2016|DCS_review_paper}.

%It has become a powerful tool in a variety of applications, including molecular spectroscopy, atmospheric sensing, biomedical diagnostics, and industrial process monitoring \cite{Baumann|2011|Characterization_methane_v3_DCS,Fleisher|2014|Real-time_detection_of_transient_free_radicals,Wei|2023|Review_DCS_chemical_application,Rieker|2014|Remote_sensing_of_greenhouse_gases,Karpov|2018|Combs_for_Biological_imaging,Miyamura|2023|Detections_of_biomolecules_DCS}. 

The two optical frequency combs used for DCS typically have repetition rates on the order of hundreds of megahertz, differing from each other by hundreds of hertz.  The beat signal between the two combs is in itself a frequency comb but now in the radio frequency (RF) domain. Each line of this comb contains information about its ``parent'' pair of optical frequencies. Detecting beat signals on a photodetector allows one to obtain information about the optical domain, enabling 
% complex spectra by using simple photodetectors as we are not dealing with optical frequencies anymore. 
direct measurement of complex spectra, for example complex molecular spectra in the near-infrared  (NIR)   range \cite{Baumann|2011|Characterization_methane_v3_DCS,Wei|2023|Review_DCS_chemical_application}.

While DCS enables high-speed measurements within a broad wavelength range, its performance is fundamentally limited by the inability to detect weak signals at each beat note frequency. In our frequency comb system, a few hundred milliwatts of total output light power is distributed across a broad spectrum (from 650\,nm to 2100\,nm), often resulting in hundreds-of-nanowatt-level power per comb tooth. Detecting these weak signals with high precision requires a detection system that combines high sensitivity, broadband coverage, and a wide dynamic range capabilities that are challenging to achieve with traditional single-element detectors.

A PIN photodiode, for example, can handle the high total power of a broad band comb light due to its large linear dynamic range and robustness. However, it lacks internal gain, which makes it poorly suited for resolving individual beat notes. When the total comb light is incident on a PIN photodiode, the weak beat note signals of individual comb teeth are masked by the noise floor, which includes contributions from shot noise, diode dark current, Johnson noise, etc. As a result, the signal-to-noise ratio (SNR) for individual beat note frequencies becomes insufficient for precise measurements \cite{Newbury|2010|SNR_and_sensitivity_of_DCS}.%\RA{???}

%\SSS{Though a PIN photodiode features a large linear dynamic range, the lack of internal gain leads to low signal-to-noise ratio (SNR), which is limited by several factors including contributions from shot noise, diode dark current, Johnson noise of the feedback or load resistor, and input current and voltage noises of amplifiers. The low SNR makes it unsuitable for precise DCS measurements. However, }
However, light-noise-limited dual-comb spectroscopy can be achieved using detectors with intrinsic amplification, such as avalanche photodiodes (APDs), photomultiplier tubes, or hybrid detectors. These detectors are capable of sensing low light levels down to a single photon. However, they cannot handle high-intensity total light. In the papers \cite{xu2024near,picque2020photon}, light-noise-limited dual-comb spectroscopy was achieved using single-photon detectors and by limiting the optical bandwidth.

 In our work, we aimed to achieve light-noise-limited detection while simultaneously measuring a broadband optical spectrum (several tens of nanometers). As proposed in Ref. \cite{Newbury|2010|SNR_and_sensitivity_of_DCS}, in a light-noise-limited condition, the use of multiple photodetectors enhances SNR. %To achieve this, we used an APD array to match the above-mentioned condition. 
 Spectrally dispersing the comb light across an array of APDs leads to several key advantages:
\begin{itemize}
%\item \textbf{Enhanced Sensitivity}: Each APD provides high internal gain, amplifying weak signals from individual beat notes above the light noise floor.
\item \textbf{Optimized Dynamic Range}: By limiting the input power to each APD by narrowing spectral slice, we avoid saturation and maintain linear operation, ensuring accurate and reliable measurements.
\item \textbf{Improved Signal-to-Noise Ratio (SNR)}: The combination of high gain and spectral filtering results in a significantly improved SNR for individual beat notes.
\item \textbf{Broadband Coverage}: The array collectively covers 52\,nm; the covarage can potentially be broadened even further.
\end{itemize}

%This approach not only overcomes the limitations of traditional detection methods but also enables high-resolution, broadband measurements that are essential for applications such as spectroscopy, optical metrology, and distance ranging. By leveraging the unique capabilities of an APD array, we demonstrate a detection system that is both highly sensitive and versatile, paving the way for advanced comb-based measurements.

% For a shot-noise-limited measurement, 
%\begin{equation}
    %SNR \approx \frac{\sqrt{\tau \dot{n}_{comb}}}{M}\,,
%\label{eq:SNR_Shot_Noise}
%\end{equation}
% \textbf{$\text{{SNR}}  \boldsymbol{\approx \sqrt{\tau n_{comb}}}\text{/M}$}
%where  $\tau$ is the acquisition time, $\dot{n}_{comb}$ is the number of detected comb photons per second 
%and $M$ is the number of the detected teeth of the comb \cite{Coddington|2016|DCS_review_paper, Newbury|2010|SNR_and_sensitivity_of_DCS}. The number of teeth appears outside of the square root because all the comb teeth rather than just the tooth of interest contribute to the noise. However, especially In the visible range, 

%Another limitation of DCS is the maximum optical range that can be acquired without aliasing when the repetition rates are fixed.
This approach opens up the possibility of performing broadband DCS  with a high SNR. However, in previous works, another limitation of DCS has been mentioned \cite{Coddington|2016|DCS_review_paper, Sugiyama|2023|DCS_visible_small_frep_alising}. The maximum optical range that can be acquired without aliasing when the repetition rates are fixed  is usually defined as
\begin{equation} 
\Delta\nu = \frac{{f_{\textrm{r}}}^2}{2 \delta f_{\textrm{r}}}\,,
\label{eq:opt_range}
\end{equation}
where $f_{\textrm{r}}$ denotes the repetition rate of one of the frequency combs and $\delta f_{\textrm{r}}$ is the difference in repetition rates between the two combs. When the recorded optical range exceeds $\Delta\nu$, the beat notes of the optical frequencies begin to alias each other, leading to ambiguity in the interpretation of the data. 
%However, in the next section, we demonstrate that the maximum optical range, under certain conditions, can be extended beyond that of Eq.\,\eqref{eq:opt_range}.
%
%\textbf{Goals of this work}
%In this paper, we demonstrate broad-band spectroscopy of samarium (Sm, atomic number $Z$=62), a highly complex atomic system, by an enhanced DCS technique designed to simultaneously probe the atoms in a broad spectral range, with high SNR. This proof-of-concept work aims to demonstrate that the system can be a valuable tool for Spectroscopy 2.0. 
%In this paper, we address these challenges by demonstrating an enhanced DCS technique with the cumulative advantages such as high SNR and broad spectral range. 
%This proof-of-concept work is used to 
This problem can be solved by selecting the repetition rates of the combs such that $k(f_\textrm{r1,2}/2)/\delta f_\textrm{r}$ is not an integer, where $k = 1,2,3,\,...$\, which introduces a relative offset between overlapping RF combs and acquiring data with high enough resolution. 

In this work, we demonstrate ambiguity-free, broadband spectroscopy ($\Delta\lambda\approx 52$\,nm) of samarium (Sm, atomic number Z=62) with a high SNR ($\approx$\,900). This enables the detection of several unreported transitions.
\begin{figure*}
\centering
\includegraphics[width=0.95\textwidth]{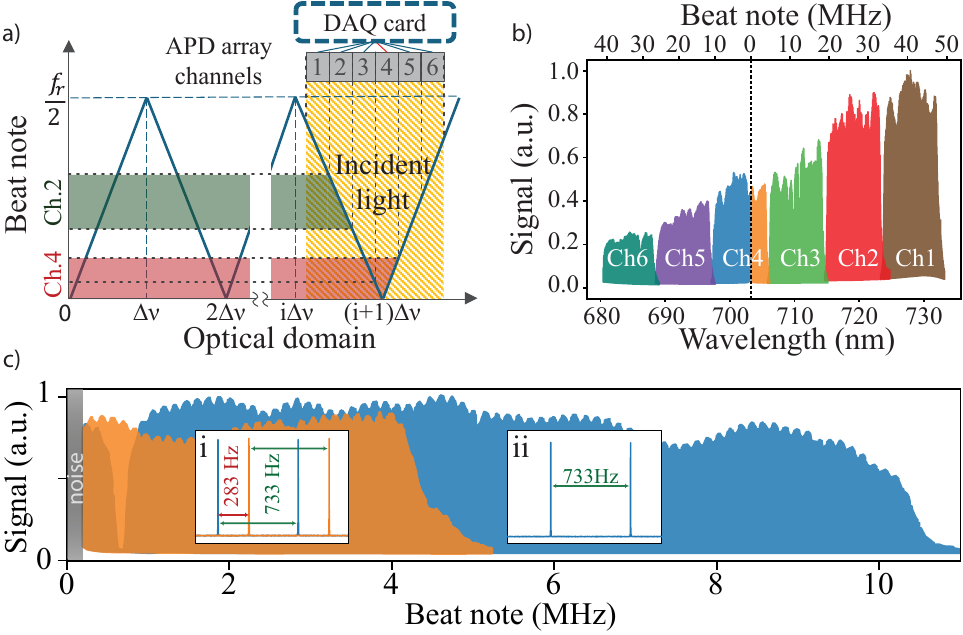}
     %\begin{subfigure}{0.48\textwidth}
     %    \includegraphics[width=\textwidth]{Opt_RF_ambiguity.pdf}
     %    \label{subfig:Opt_RF_ambiguity}
     %\end{subfigure}
     %\hfill
     %\begin{subfigure}{0.48\textwidth}
     %    \includegraphics[width=0.98\textwidth]{full_spec_nm_rf_1.pdf}
     %    \label{subfig:full_spectrum_channels}
     %\end{subfigure}
    
     %\vspace{0.5cm} % Adds space between the figures

    % % Second subfigure (Bottom)
     %\begin{subfigure}{\textwidth} % Full width
     %    \centering
     %    \textbf{(a)}\\
     %    \includegraphics[width=0.95\textwidth]%{ambiguity_full_spectrum2.pdf}
     %    \label{subfig:ambiguity_full_spectrum}
     %\end{subfigure}

\caption{ Mapping of the incident light frequencies to the RF frequencies:
    a) The general mapping method, with an example of ambiguous and unambiguous data. Channel 2 (unambiguous data) and channel 4 (ambiguous data) are labeled as Ch.\,2 and Ch.\,4, respectively. 
    b) The real data from six channels illustrates the signal along with its recorded RF frequencies and corresponding wavelengths. This data represents Sm spectrum at 1040\,$^\circ$C.
    c) The normalized recorded spectrum on channel 4, where ambiguity is expected (for $\delta f_{\textrm{r}}\approx 733$\,Hz). The blue and orange data represent spectra from different unambiguous regions. These components were separated during the data analysis process. As illustrated in the two smaller zoomed-in figures when the two spectra overlap (i), we observe two sets of teeth with a relative offset of 283 Hz, whereas in case of no ambiguity, we see only one set of teeth with a 733 Hz spacing (ii). Moreover in the overlapping region, a dip in the blue spectra corresponds to a strong absorption line of samarium.
    %also in the overlapping region, a strong absorption line of Sm is visible in the blue spectrum. When there is no ambiguity, we see only one set of teeth with a 733 Hz spacing (ii).
    %Two sets of combs, having a 283 Hz offset from each other, were aliasing expected, and $\delta f_{\textrm{r}}$=733\,Hz.
}
\label{fig:mapping}
%Channel 2 and Ch.\,4 represent frequencies recorded with  photodetectors Ch.\,2 (unambiguous data) and Ch.\,4 (ambiguous data), respectively.}
\end{figure*}
\section{Ambiguity in frequency conversion} \label{Ambiguity}
The beat signal between the two combs, acquired by a photodetector, contains radio frequencies, and to retrieve the corresponding optical frequencies, we must convert this RF spectrum back into optical domain. A visual representation of this frequency mapping, Fig.\,\ref{fig:mapping}(a), can help to understand the conversion process. To ensure that only the beat note between neighboring optical lines is acquired, the maximum frequency in the RF domain (the vertical axis) should be limited to $\le f_r/2$. %To achieve this, a low-pass filter should be applied to the signal.
The values i$\Delta\nu$, (where i = 0,1,2,... and $\Delta\nu$ is defined in Eq.\,\eqref{eq:opt_range}) delimit $i$th unambiguity region on the horizontal axis. The black vertical dashed lines denote the optical frequency ranges captured by each photodiode, whereas the horizontal dashed lines represent the RF frequencies of the corresponding recorded signals. 
To convert the RF spectrum we should start with the general representation of
the optical frequencies of the comb:
\begin{equation} 
\begin{split}
\nu_{{\textrm{1}n}} = f_{{\textrm{CEO1}}} + nf_{{\textrm{r1}}}\,,\\
\nu_{{\textrm{2}m}} = f_{{\textrm{CEO2}}} + mf_{{\textrm{r2}}}\,,
\end{split}
\label{eq:comb}
\end{equation}
where $n$ and $m$ represent the tooth numbers of corresponding optical frequencies $\nu_{1n}$ and $\nu_{2m}$ of the first and second comb, respectively, and $f_{{\textrm{CEO1,2}}}$ is the carrier-envelope offset. In our case, we lock the system such that $f_{{\textrm{CEO1}}} = f_{{\textrm{CEO2}}}$.

Since both combs contribute equally to each beat note, it is necessary to select one of them to serve as a reference for our calculation frame. We will choose the comb with the smaller repetition rate (this is an arbitrary choice); let us consider $f_{{\textrm{r1}}}\,<\,f_{{\textrm{r2}}}$. The beat note ($f_{{\textrm{RF}}}$) between pair of teeth is:
\begin{equation} 
f_{{\textrm{RF}}}\,=\,|\nu_{1n}\,-\,\nu_{2m}|\,= |nf_{{\textrm{r1}}} - mf_{{\textrm{r2}}}|\,.
\label{eq:RF}
\end{equation}
The relationship between $n$ and $m$ for neighboring teeth depends on the number of $f_{\text{r1}} / 2$ border crossings ($j$) up to that region, such that $n = m - j$. By comparing these RF frequencies with the recorded ones, we can reconstruct the corresponding optical frequencies. However, to use this technique, we must ensure that the $f_{\text{RF}}$ from different regions are unambiguously distinguished. An ambiguity would occur when $k(f_\textrm{r1}/2)/\delta f_\textrm{r}$ is an integer, where $k = 1,2,3,\,...$\,.
As shown in Fig.\,\ref{fig:mapping}(a), channel 4 contains signals from two different regions and hence exhibits ambiguity, while other channels such as channel 2 remain unambiguous. In an experiment, an additional complication may arise due to cross-talk between photodiodes.

%In practice, it is impossible to prepare an experiment in such a way as to have no ambiguity in every diode, as optical filters are not perfect and cannot fully isolate the spectrum. Consequently, there will always be an overlaying spectrum near $0$ and $\frac{f_r}{2}$.

We propose to resolve the ambiguity in the following manner: when $k(f_\textrm{r1}/2)/\delta f_\textrm{r}$ is not an integer, RF combs from different regions will exhibit a relative offset. This offset can be observed if the acquisition time is longer than the inverse of relative offset or, in other words, spectral resolution is better than relative offset. The relative offset can be determined by comparing simulated RF combs from different regions of unambiguity with each other.
In Fig.\,\ref{fig:mapping}(b), we present the data acquired with our system. The expected ambiguity in the spectrum as recorded in channel 4 is shown in Fig.\ref{fig:mapping}(c). In this case, $\delta f_{\textrm{r}}$ $\approx$\,733.04\,Hz, so if aliasing were present, there would be only teeth separated by $\delta f_{\textrm{r}}$. However, we observe two sets of comb lines with a relative offset of 283\,Hz in the overlapping region of the spectra as shown in the inset (i) of Fig.\,\ref{fig:mapping}(c). ``Unfolding'' the data as described above allows us to attribute the spectral peaks to their correct optical frequencies.

One can resolve this ambiguity by aligning the system so that the crossing occurs exactly between two channels. However, when multiple crossings occur across the spectrum, it becomes difficult to tune the system in a way that ensures none of the photodiodes exhibit ambiguity. Instead, addressing these ambiguities during post-analysis using the method described above is significantly easier and more practical.

%two RF combs with a relative offset are observed. 
%This result was recorded with our dual-comb system as proof of the effect. 
%\begin{figure}[h!]
%\centering
%\includegraphics[width=6.4cm, height=4.7cm]{ambiguity.pdf}
%\caption{\label{fig:ambiguity}Two sets of combs having a X Hz offset from each other were expected to cause aliasing.}
%\end{figure}

%This approach ensures that the data are completely free from aliasing, allowing the full spectrum to be accurately unfolded from the recorded data.% It also eliminates the need for prior knowledge of the optical frequencies, as all information can be analyzed post-acquisition. Each RF comb offset is linked to the unambiguous range, making the repetition rates and CEO frequencies of combs the only initial information required.
\section{Experimental setup} \label{Experimental setup}
%\textbf{The basics: two Menlo combs}
%

\begin{figure}
\centering
\includegraphics[width = \linewidth]{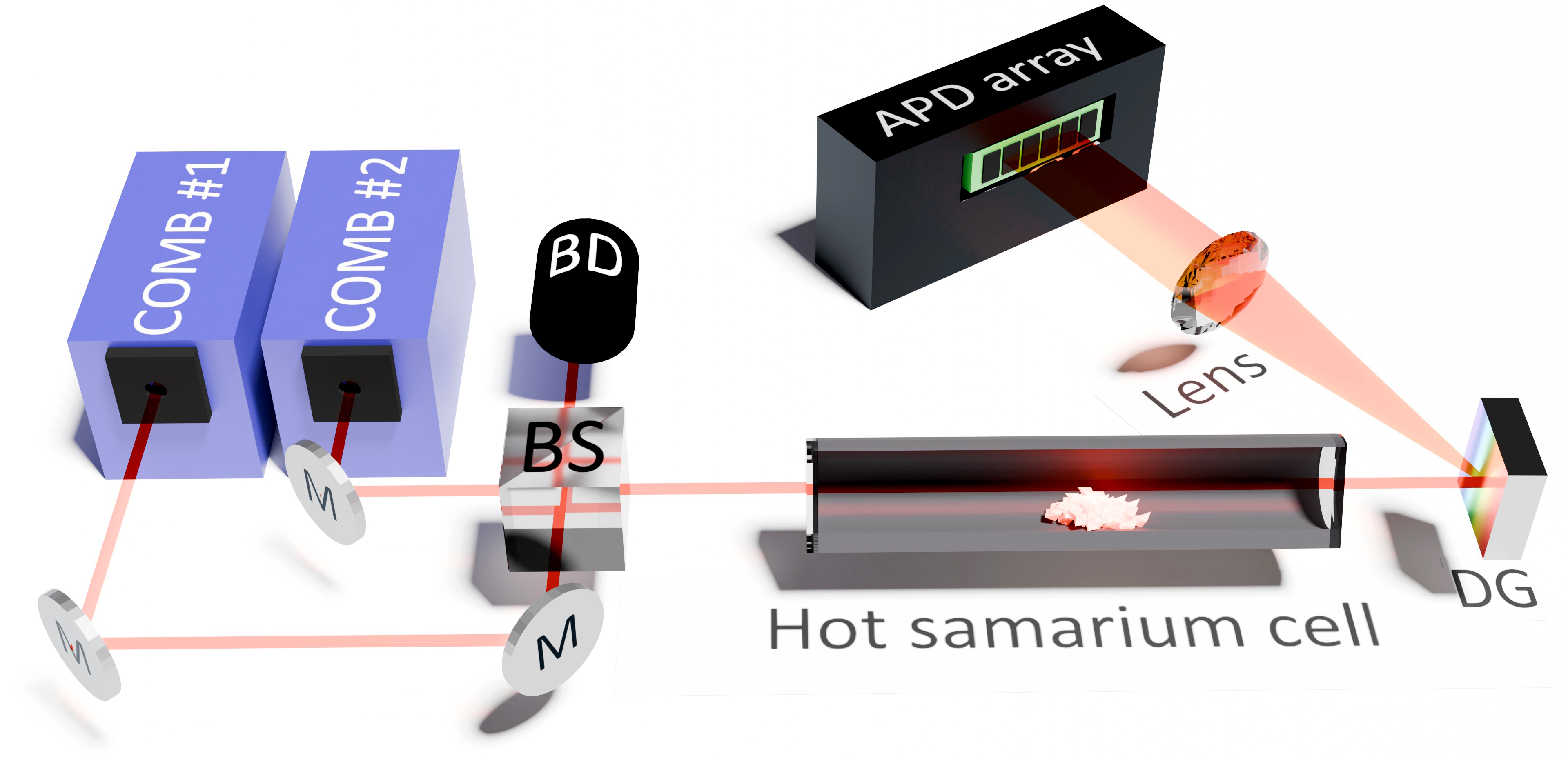}
\caption{Schematic of the experimental setup, where  M\,:\,mirror, BS\,:\,beam splitter, BD\,:\,beam dump and DG\,:\,diffraction grating. %\OT{Do we used lenses around grating? Maybe it is nice to show them as well as say that "good focused light on the array" is not good because of the gaps between channels. Maybe we can add that we did 2 or 3 passes to get a sufficient optical pass.}
}
\label{fig:setup}
\end{figure}

The experimental setup (Fig.\,\ref{fig:setup})
incorporates a pair of MenloSystems FC1500-250-ULN combs locked to a low phase noise, narrow linewidth ($<1$\,kHz) 1542\,nm laser diode (RIO Planex). Then, the repetition rate of each comb follows the laser diode frequency and is stabilized to an FS725 rubidium frequency standard (Stanford Research Systems) via the laser diode. The system generates femtosecond pulses with repetition rates $\approx$\,250\,MHz ($f_{\textrm{r1}} \approx 250000240.75\,\textrm{Hz}$ and $f_{\textrm{r2}} \approx 250000973.79\,\textrm{Hz}$) and covers a spectral range from 650\,nm to 2100\,nm. The outputs of the two combs are combined using a beam splitter and passed through a samarium (Sm) cell maintained at a temperature of $\approx 1040^\circ$C to produce a sufficient density of Sm atoms at a buffer gas (argon) pressure of 330\,torr. 
At the temperature and pressure used, the combined Doppler and collisionally broadened linewidth of atomic transitions (on the order of a gigahertz) is sufficiently large, so that several comb teeth ($\approx$250\,MHz separation defined by the pulse repetition rate) fall under the line profile.  
The combined light beam 
%passing the vapor cell
is spectrally dispersed with a grating onto the avalanche photodiode (APD) array. 

The APD array used in this experiment is Hamamatsu S15249. It consists of 1\,x\,16 diodes with separate output channels of uniform gain. In this work, only six of them were used as a proof of principle.
% \OT{more or less. According to specs $\pm 10 \% $max which could be compensated during post-processing}. 
The array has a spectral response range of 350\,-\,1000\,nm, with peak sensitivity at 620\,nm. We equip each APD channel with a dedicated transimpedance amplifier with a 125\,MHz bandwidth, placed as close as possible to the APD array chip, and allowing simultaneous operation across all channels. The common cathode was powered by a linear voltage-stabilized power supply, ensuring low-noise performance, particularly at high frequencies ($>50$\,Hz). The APD of bias voltage was adjusted to optimize the APD gain and achieve the best possible SNR. This choice is motivated by the low output power of the laser in the visible range, where it is critical to ensure that the background noise of the photodetector remains smaller than the laser light noise.

%As we saw, acquiring high-resolution data is essential to effectively addressing the ambiguity problem. To achieve this, we set the acquisition time to ${\tau \approx 1 \,\text{s}}$, providing a frequency resolution of $\approx 1 \,\text{Hz}$. This resolution is adequate to resolve the offsets caused by crossing the unambiguous region.
In this work, we used the same acquisition system as described for ``Apparatus A'' in \cite{Tretiak|2022|DarkMatter}. 
Specifically, we used a two-channel digitizer (Spectrum M4i.4420-x8) with a 250\,MHz sampling rate and 16-bit resolution. The digitizers clock frequency was locked to the same rubidium frequency standard as the combs.  

The system captures spectra up to 125\,MHz (the Nyquist frequency) with a resolution of 0.93\,Hz. Data were recorded in chunks of $2^{28}$ samples per channel ($\approx 1.073$\,s of data) and Fourier-transformed on a graphics card in parallel, ensuring continuous acquisition without interruptions. For simplicity, we call it a ``one-second" data record.  In this work, we did not apply any special window function to the data (a rectangular window was used). The resulting spectra were averaged using the central processor.  For details on the data acquisition system, refer to the supplementary materials \cite{Tretiak|2022|DarkMatter}.

The system imposes no restrictions on the number of spectra that can be averaged. In our experiment, we averaged 1000 spectra per measurement ($\approx 17$~min). Data from both channels were collected simultaneously. Since the system operates in parallel and each channel functions independently, the number of computers with digitizers can be scaled to enable simultaneous recording across any number of channels. The independent systems can be synchronized using an external trigger.  
 
The high spectral resolution reveals a scalloping pattern between the acquisition system and the RF comb. Various methods can mitigate these artifacts during post-analysis. Our approach involves detecting the peaks of the teeth in the recorded spectrum and redistributing the energy of neighboring bins of the tooth.

\section{Results} \label{Results}

\begin{figure*}
\centering
\includegraphics[width = \linewidth]{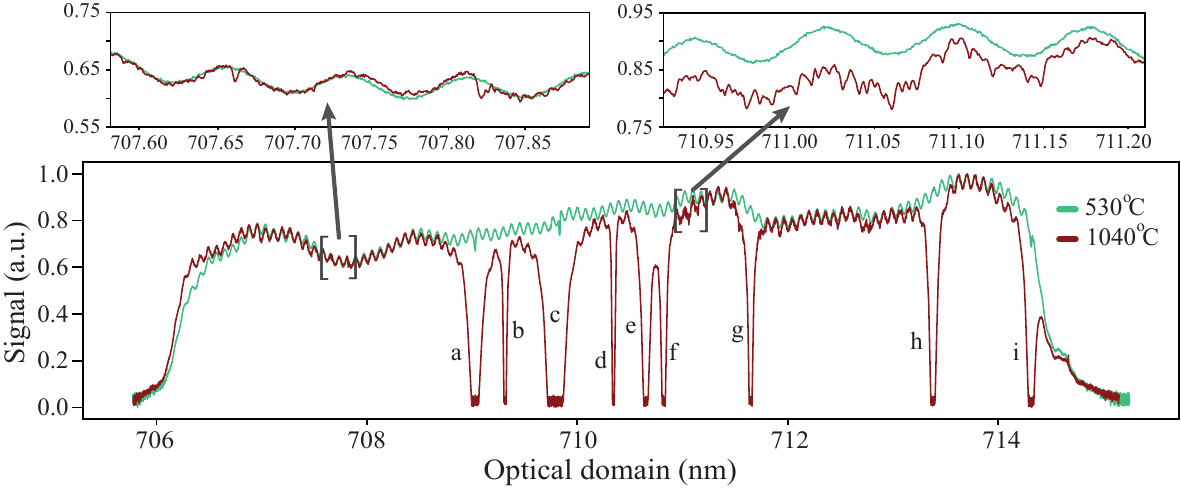}
\caption{Absorption spectrum of Sm vapor at various temperatures, obtained by averaging 1000 spectra of one-second records on channel 3, resulting in an SNR of 1100.   The smaller plots show zoomed-in regions where several unknown lines have been detected. The numerical values of the wavelengths for strong transitions are: a\,-\,709.030\,nm, b\,-\,709.314\,nm, c\,-\,709.828\,nm, d\,-\,710.343\,nm, e\,-\,710.653\,nm, f\,-\,710.823\,nm, g\,-\,711.648\,nm, h\,-\,713.383\,nm, and i\,-\,714.316\,nm}

\label{fig:710_spectrum_alp}
\end{figure*}

The incident light had an optical bandwidth of $\approx 9 \, \text{nm}$ on each photodiode. As our acquisition system has only two parallel data channels, we performed three sets of two-channel acquisitions (parallel channels: 1\&2\,$\mid$\,3\&4\,$\mid$\,5\&6), covering a total optical range of $\approx 52 \, \text{nm}$ (680\,nm to 732\,nm) shown in Fig.\ref{fig:mapping}(b). To maintain a high SNR we averaged 1000 one-second recordings. The SNR varies across the spectrum due to intensity change, as the recorded spectrum lies on the edge of the emission spectrum of the combs. The average SNR is $\approx 900$, the SNR was calculated as follows: 
%in the frequency domain by dividing the peak minus the baseline of each RF comb line by the standard deviation of noise from the baseline in the region between two peaks.\begin{equation} 
\begin{equation}
    SNR = \frac{\sqrt{Tooth\ Height^2 - Baseline^2}}{\sigma}\,,
\label{eq:SNR_cal}    
\end{equation}

where $\sigma$ is the standard deviation of the points between two teeth.
This approach achieves a DCS figure of merit of $M \times \text{SNR} / \sqrt{\tau} = 3.5 \times 10^6 \, \text{Hz}^{1/2}$, and constrained only by the number of the used elements of the photodetector array. 
In Fig.\,\ref{fig:710_spectrum_alp}, the most interesting spectrum of the six channels is presented, showing several absorption lines. The high-resolution plots of this spectrum and the remaining channels can be found in the supplementary materials. We compared measurements taken at different cell temperatures, $\approx 530^\circ$C and $\approx 1040^\circ$C, to analyze spectral variations at low and high samarium concentrations. These data can also be used to normalize the etalon effect caused by optical elements, which appear as oscillations in the spectrum. 
%Several unknown lines were detected in this spectral range, along with known ones. Data are compared with an existing dataset \cite{Ferrara|2023|Spectral_lines,Martin|1978|Energy_levels,Meggers|1975|Spectral_lines}. Two zoomed-in subplots are provided to illustrate the difference between dense and sparse spectral regions, ensuring that spectral transitions are not mistaken for noise. These unknown transitions require further investigation for characterization.

%\RA{Numerous unknown transition lines, along with known ones, were detected in this spectral range. The data are compared with an existing dataset \cite{Ferrara|2023|Spectral_lines,Martin|1978|Energy_levels,Meggers|1975|Spectral_lines}. 
When compared with the existing dataset \cite{Ferrara|2023|Spectral_lines,Martin|1978|Energy_levels,Meggers|1975|Spectral_lines}, numerous previously unknown lines were detected.
Two zoomed-in subplots illustrate spectral regions with sparse and dense transition lines. To ensure that dense spectral regions are not misinterpreted as system noise, the data at $\approx 530^\circ$C is provided, demonstrating that the noise level remains consistent across different spectral regions. The detected unknown transitions require further investigation for proper characterization.
%Table \ref{tab:transitions} presents the  lines that are labeled in the plot, all of which correspond to previously identified transitions.
%\begin{table}[h]

%\caption{\label{tab:transitions} Table of the labeled transitions in Sm\,I.
%}
%\centering
%\begin{tabular}{cc}  
    %\hline
    %\textbf{Label} & \textbf{Wavelength (nm)} \\ 
    %\hline
    %a & 709.030 \\
    %b & 709.314 \\
    %c & 709.828 \\
    %d & 710.343 \\
    %e & 710.653 \\
    %f & 710.823 \\
    %g & 711.648 \\
    %h & 713.383 \\
    %i & 714.316 \\
    %\hline
%\end{tabular}
%\end{table}
\section{Conclusion} \label{Conclusion}
We present a successful enhancement of the DCS technique by incorporating a photodetector array, effectively addressing the trade-off between SNR and optical bandwidth. It also demonstrates the capability of this technique to acquire a broad spectrum without ambiguity, by introducing a relative offset to RF comb tooth sets from different unambiguity regions. By acquiring data with high spectral resolution, we were able to resolve and reconstruct the unambiguous spectrum. Furthermore, this approach enabled the detection of numerous previously unreported transition lines, highlighting its potential for discovering new spectral features.

The ability of this system to acquire spectra with both broad bandwidth and high SNR makes it a valuable tool for Spectroscopy 2.0. The next step of this project is to design and implement experiments at high magnetic field environments, up to the 100T\,\cite{Hervieu|2015|43T_Magnet, Beard|2018|100T_Magnet}, to explore the effects of strong fields on atomic transitions. 

%As we have presented, samarium exhibits a highly complex absorption spectrum with closely spaced transitions, making human interpretation particularly challenging, especially when dealing with large datasets, even more so when strong magnetic fields are applied. However, with rapid advancements in computational power and machine learning, spectral analysis is undergoing a transformation. These technologies enable automated classification, anomaly detection, and theoretical predictions, which will significantly enhance our ability to analyze and interpret complex atomic spectra.

\section{Acknowledgement}
The authors thank Nathalie Picqu\'e, Jason Stalnaker, Arman Cing\"oz, and Mikhail Kozlov for helpful discussions.
This work was supported by the European Commission Horizon Europe Framework Program under the Research and Innovation Action MUQUABIS GA no. 101070546.

\section{Data availability}
The data supporting the plots in this article and the supplementary plots are available from the corresponding author upon request.

\bibliography{ref}

\end{document}